\def\arxiv{}

\ifdefined\thesis
    \documentclass[reprint,NumberedRefs]{JASA_blank}
    \def\showfigures{}
\else
\ifdefined\reprint
    \documentclass[reprint,NumberedRefs]{JASA}
    \def\showfigures{}
\else
\ifdefined\reviewers
    \documentclass[preprint,NumberedRefs,TurnOnLineNumbers,trackchanges]{JASA}
    \def\showfigures{}
\else
\ifdefined\submission
    \documentclass[preprint,NumberedRefs,TurnOnLineNumbers]{JASA}
    \def\showfigures{}
\else
\ifdefined\arxiv
   \documentclass[twocolumn,amsmath,amssymb,reprint,aip,floatfix]{revtex4-1}
    \def\showfigures{}
    \usepackage{graphicx}
    \usepackage{bm}
    \usepackage{amsfonts}
    \usepackage{latexsym}%
    \usepackage{xcolor}
    \usepackage[bookmarks=false,         
         pdfnewwindow=true,      
         colorlinks=true,    
    final=true
     ]{hyperref}
    \newdimen\reprintcolumnwidth
    \global\reprintcolumnwidth=246pt
    \def\dodoi#1{doi: \href{https://doi.org/#1}{\nolinkurl{#1}}}
    \def\dourl#1{\href{http://#1}{\nolinkurl{#1}}}
\else 
    \documentclass[preprint,NumberedRefs]{JASA}
    \def\appendcaptions{}
\fi
\fi
\fi
\fi

\graphicspath{{.}}

\usepackage{subfigure}
\usepackage{multirow}
\renewcommand{\vec}[1]{\boldsymbol{\mathbf{#1}}}  
\newcommand{\mat}[1]{\boldsymbol{\mathbf{\MakeUppercase{#1}}}}  

\newcommand{\diff}{\text{d}}
\newcommand{\netp}{p_{\mathbf{\bullet}}}
\newcommand{\netu}{\mathbf{u}_{\mathbf{\bullet}}}
\newcommand{\netI}{\mathbf{I}_{\mathbf{\bullet}}}

\newcommand{\R}{\mathbb{R}}

\newcommand{\E}{\mathbb{E}}

\newcommand{\rr}{\vec{r}}

\renewcommand{\P}{\vec{P}}

\DeclareMathOperator*{\argmin}{arg\,min}

\begin{document}

\title[PINNs for room impulse response reconstruction]{Room impulse response reconstruction with physics-informed deep learning}


\author{Xenofon Karakonstantis}


\affiliation{Acoustic Technology, Dept. of Electrical \& Photonics Engineering, Technical University of Denmark}
\author{Diego Caviedes-Nozal}
\affiliation{Audio Research, GN Audio A/S \& Jabra}
\author{Antoine Richard}
\affiliation{Odeon A/S}
\author{Efren Fernandez-Grande}
\affiliation{Acoustic Technology, Dept. of Electrical \& Photonics Engineering, Technical University of Denmark}




\email{efgr@dtu.dk}



\date{\today} 

\begin{abstract}
A method is presented for estimating and reconstructing the sound field within a room using physics-informed neural networks. By incorporating a limited set of experimental room impulse responses as training data, this approach combines neural network processing capabilities with the underlying physics of sound propagation, as articulated by the wave equation. The network's ability to estimate particle velocity and intensity, in addition to sound pressure, demonstrates its capacity to represent the flow of acoustic energy and completely characterise the sound field with only a few measurements. Additionally, an investigation into the potential of this network as a tool for improving acoustic simulations is conducted. This is due to its profficiency in offering grid-free sound field mappings with minimal inference time. Furthermore, a study is carried out which encompasses comparative analyses against current approaches for sound field reconstruction. Specifically, the proposed approach is evaluated against both data-driven techniques and elementary wave-based regression methods. The results demonstrate that the physics-informed neural network stands out when reconstructing the early part of the room impulse response, while simultaneously allowing for complete sound field characterisation in the time domain.
\end{abstract}


\maketitle



\section{Introduction}
\noindent
In the search for immersive virtual experiences and personalized acoustic environments, the accurate reconstruction of sound fields within complex spaces has become a paramount challenge in various applications of audio and acoustics. Sound field reconstruction, as the process of interpolating or extrapolating sound fields in a given environment using virtual sources (e.g., basis functions) and sound field measurements, plays a crucial role in achieving realistic acoustic conditions. Sound field reconstruction techniques (understood as interpolating and extrapolating a sound field from a limited set of spatio-temporal data) find a broad range of applications, including sound field analysis for acoustic environment design and optimisation,\cite{de1999wave, verburg2018reconstruction, nolan2019experimental} spatial audio implementation in virtual, augmented and mixed reality (AR/VR/XR) systems,\cite{spors2013spatial, kirkeby1996local, ahrens2019auralization} and sound field control,\cite{betlehem2015personal, moller2016sound, coleman2014acoustic} where active manipulation of the sound field enables specific acoustic goals to be achieved. While traditional signal processing approaches have been prevalent, the increasing demand for precision and realism drives the pursuit of innovative computational techniques that integrate physics with state-of-the-art technology.

%
A widely employed technique for sound field reconstruction involves the use of basis function expansions, which enable the representation of the sound field as a linear combination of predetermined basis functions. 
The selection of specific basis functions is driven by the desired application and trade-offs between accuracy and computational complexity. In recent literature, several approaches have gained popularity, such as wave superposition, \cite{verburg2018reconstruction, caviedes2021gaussian, zea2019compressed} room mode decomposition, \cite{haneda1999common, mignot2013low, das2021room} spherical harmonics decomposition, \cite{wang2017compressive, nolan2019experimental} and kernel methods. \cite{ueno2018kernel, caviedes2021gaussian, duran2022reconstruction}

One of the main tasks/objectives in sound field reconstruction is to accurately represent the sound field at locations where no measurements are available. This requires the use of interpolation or extrapolation techniques to estimate the sound field at these locations based on the available measurements. A number of methods have been proposed to address this challenge, including regularisation techniques that impose sparsity constraints on the reconstructed sound field, \cite{verburg2018reconstruction,antonello2017room, caviedes2023spatio} as well as machine learning approaches that use data-driven models to predict the sound field at unobserved locations. \cite{lluis2020sound,karakonstantis2021sound,fernandez2023generative} More recently, models that attempt to physically interpret the heuristic decomposition of room impulse responses (RIRs) \cite{fernandezreconstruction, duran2022reconstruction} have seen success in this task.
%

Deep learning methodologies offer a promising approach for sound field reconstruction, drawing significant interest in this domain. Convolutional neural networks (CNNs) and generative adversarial networks (GANs) \cite{goodfellow2016deep} are prominent models that excel at capturing complex relationships between input data (sound field measurements or samples) and the corresponding output (reconstructed sound field). This adaptability suits sound field reconstruction tasks, which often involve intricate associations and may require prior knowledge of room properties. For instance, Lluis et al. \cite{lluis2020sound} proposed a U-net architecture trained with the magnitude of sound fields obtained using Greens function in rectangular enclosures. Similarly, Karakonstantis and Fernandez-Grande \cite{karakonstantis2021sound} and Fernandez-Grande et al. \cite{fernandez2023generative} employed a GAN-based approach to reconstruct sound fields from limited measurements, demonstrating superior performance over classical techniques.

One promising approach that has recently gained attention is the use of physics-informed neural networks (PINNs). \cite{PINNoriginal, borrel2021PINN, shigemi2022physics} PINNs are neural networks that are trained to solve partial differential equations (PDEs), such as the wave equation, \cite{borrel2021PINN, rasht2022physics} which govern the propagation of sound waves in a given environment. By incorporating physical principles into the training process, PINNs can learn to accurately reconstruct sound fields via capturing the underlying physics of the problem. This makes them a promising candidate model for sound field reconstruction, as they can potentially offer improved accuracy and computational efficiency compared to traditional methods. Furthermore, as they are function approximators,\cite{hornik1989multilayer} they allow for a continuous representation of the sound field which they are posed to reconstruct (grid-free mapping). 

In this paper, we explore two contexts for employing PINNs. The first involves experimental measurements in a room, while the second deals with a numerical simulation of the sound field in the same room. Using PINNs on the simulated sound field allows us to evaluate their potential as enhancements to simulations, enabling gridless representation and real-time auralisation from any desired position within the enclosure. This opens up many possibilities for immersive auditory experiences. On the other hand, the experimental dataset showcases the PINN's adaptability to realistic scenarios, including limited measurements, measurement noise and other complexities. Despite these challenges, the PINN method enables a comprehensive characterisation of the experimental sound field quantities. By investigating PINNs on these two sound fields, we gain insights into their benefits for real-time auralisation and their effectiveness in addressing challenging and noisy real-world acoustic environments.
%
\section{Theoretical Background}\label{sec:theory}
\subsection{Surrogate neural networks for solving partial differential equations}

Surrogate neural networks have gained popularity as a powerful tool for solving PDEs, thanks to their ability to approximate complex functions and handle high-dimensional input spaces. Surrogate networks typically receive a combination of Cartesian coordinates $\rr$ and an instance of time $t$ as input, and return the value of the approximated eigenfunction $\Phi(\mathbf{r})$ of the PDE at the coordinate in a predetermined domain $\Omega_{m}$. These neural networks are trained to fit the forward problem so that 
\begin{align}
\text { find } \Phi(t, \rr) \quad &\text { s.t. } \mathcal{C}_{k}(\mathbf{\chi}(t, \rr), \Phi(t, \rr), \nabla \Phi(t, \rr), \ldots)=0, \\ &\;\forall \;\mathbf{r} \in \Omega_{k} \And t \in \mathbb{R}^{+}, k=1, \ldots, K \nonumber,
\end{align}
where $\mathcal{C}_{k}$ refers to a set of $K$ constraints, most times in the form of the PDEs themselves, which relate the eigenfunctions $\Phi$ to any arbitrary forcing function $\mathbf{\chi}$. These networks are trained with the sum of the various constraints applied,\cite{PINNoriginal,cuomo2022scientific} as denoted by 
\begin{equation}
\mathcal{L}=\int_{\Omega} \sum_{k=1}^{K} \mathbf{1}_{\Omega_{k}}(\mathbf{r},t)\left\|\mathcal{C}_{k}(\mathbf{\chi}(\mathbf{r},t), \Phi(\mathbf{r},t), \nabla \Phi(\mathbf{r},t), \ldots)\right\| d \mathbf{r}  dt,
\end{equation}
where $\mathbf{1}_{\Omega_{k}}(\mathbf{r}) = 1$ indicates that $\mathbf{r} \in \Omega_{k}$ and is zero otherwise. This can be understood as a binary mask that specifies which constraints $\mathcal{C}_{k}(.)$ are applied to each point within the domain $\Omega$. Since $\Omega$ is a continuous space, we sample the points $\mathbf{r}$ within the bounds defined by $\Omega$ at each iteration of the neural network training process.
\subsection{Sound field reconstruction}
\noindent
Linear sound fields excited by a source can be described by the inhomogeneous wave equation as
\cite{jacobsen2013fundamentals}
\begin{equation}\label{eq:wave_eq}
    \nabla^2p(t, \rr) - \frac{1}{c^2}\frac{\partial^2p(t, \rr)}{\partial t^2} = \chi(t, \rr_s).
\end{equation}
This equation describes the sound pressure $p(t, \rr)$ as a function of time $t$ and space $\rr$ at a speed of sound $c$, with an arbitrary excitation term $\chi(t, \rr_s)$ representing the source located at position $\rr_s \subset \rr$. Considering an incomplete acquisition of the sound field, where only a limited set of measurements are available, denoted as $\tilde{p}(t_n, \rr_{m})$, at positions $\rr_{m}$ and time $t_{n}$ with $\rr_{m} \subset \rr$ and $t_n \subset t$. Our goal is to find a function that can accurately express the underlying pressure field $p(t, \rr)$ based on only a few noisy observations $\tilde{p}(t_n, \rr_{m})$.

One approach to estimating the pressure field is by expanding it onto a set of basis functions and regressing the unknown weights of those functions. This enables the approximation of the pressure $p(t, \rr)$ numerically through the linear combination of basis functions as \cite{costabel2004time}
\begin{equation}\label{eq:convolution}
p(t, \rr) \simeq \sum_{l=1}^{L} \varphi_{l}(t, \rr) * \alpha_{l}(t),
\end{equation}
where $\varphi_{l}(t, \rr)$ refers to the basis function, $\alpha_{l}(t)$ represents the unknown coefficients to be obtained by solving the inverse problem given measurements $\tilde{p}(t_n, \rr_{m})$. The operator $*$ denotes linear convolution, and $L$ is the number of basis functions. Similarly, this linear combination of basis functions can be expressed in the frequency domain for an angular frequency $\omega$ as
\begin{equation}\label{eq:freq_domain_basis}
p(\omega, \rr) \simeq \sum_{l=1}^{L} \psi_{l}(\omega, \rr) \cdot \beta_{l}(\omega),
\end{equation}
where $\psi_{l}(\omega, \rr)$ represents the analytic basis functions used to express the sound field harmonically. This last approach enables us to project the sound field onto these basis functions, providing a solution to the homogeneous Helmholtz equation \cite{fourier_acoustics} or the Fourier transform of the left-hand side of Eq.\,\eqref{eq:wave_eq}.

Recently, deep learning methods have been applied to address the challenge of reconstructing missing pressure values within a discretised domain.\cite{pezzoli2022deep, lluis2020sound} The objective is to obtain the pressure field, denoted as $\P \in \R^{N\times M^{}}$, which represents discrete pressure values at $N$ time samples and $M^{}$ discrete positions. This is achieved through the forward pass of a neural network $\mathcal{NN}(\cdot)$, taking the measured pressure $\tilde{\P} \in \R^{N \times M}$, acquired at $M$ measurement positions, as its input. This can be described by 
\begin{equation}
\P = \mathcal{NN}(\tilde{\P}),
\end{equation}
where the neural network attempts to reconstruct the pressure field $\P$ by leveraging the spatial correlation within the measured pressure data $\tilde{\P}$.

However, most of the existing deep learning models lack certain implicit biases, such as wave propagation characteristics, which can substantially enhance the reconstruction performance. To address this limitation, we propose an alternative approach using a PINN that models the spatio-temporal function of pressure in a continuous domain. The proposed PINN is able to utilise both the measurement data and the implicit bias associated with wave propagation.
%

%
%
\subsection{Physics informed neural network for sound field estimation}
\noindent
Typically, a PINN would be represented by a multi-layer perceptron (MLP). Given an input $[x,y,t]^{\intercal} = [\rr^{\intercal},t]^{\intercal}$ of Cartesian coordinates $x, y$ and time $t$, the pressure $p(t, \rr)$ is given by
\begin{equation}
\begin{split}
      p(t,\rr) &= \mathbf{W}^d \cdot \sigma\Big(\mathbf{W}^{d-1} \cdot \sigma\Big(\dots \cdot \sigma\Big(\mathbf{W}^1 \cdot [\rr^{\intercal},t]^{\intercal} \Big) \Big)\quad\Big).
\end{split}
\end{equation}
Here $d$ is the number of hidden layers, $\mathbf {W}^1, \dots, \mathbf{W}^d$ are the weight matrices and the bias vectors $\mathbf{b}^1, \dots, \mathbf{b}^d$ have been omitted for brevity. The function $\sigma$ is the non-linear activation function. As MLPs are susceptible to stiffness of gradient flow dynamics, and can often fall short of correctly approximating the Hessian of hyperbolic equations such as Eq.\,\eqref{eq:wave_eq}, we adopt a specialised neural network architecture, also known as modified multi-layer perceptron and introduced by Wang et al. \cite{wang2020gradient_pathologies} 

This modified MLP (mMLP) architecture is inspired by transformer attention mechanisms, \cite{vaswani2017attention} to enhance the effectiveness of PINNs. The mMLP incorporates input variables $[\rr^{\intercal},t]^{\intercal}$ into network hidden states. Initially, two encoders, $U$ and $V$, transform the inputs into a feature space. These encoded inputs are then integrated into each hidden layer of a standard MLP using element-wise multiplication. For the input $\vec{x} = [\rr^{\intercal},t]^{\intercal}$, the encoders are defined as
\begin{equation}
U=\sigma\left(\mathbf{W}^U \vec{x}\right), \quad V=\sigma\left(\mathbf{W}^V  \vec{x} \right).
\end{equation}
Hence, each forward pass is defined as
\begin{equation}
\begin{gathered}
Z^l(\mathbf{x})= \sigma \left( \mathbf{W}^l Z^{l-1}(\mathbf{x}) \right) , \quad \text { for } l \in\{1,2, \ldots, d \}\\
Z^l(\mathbf{x})=\left(1-Z^l(\mathbf{x})\right) \odot U+Z^l(\mathbf{x}) \odot V \\
\netp(t, \rr) =\sigma\left( \mathbf{W}^d Z^d(\mathbf{x})\right)
\end{gathered}
\end{equation}
where $\odot$ denotes element-wise multiplication between the output $Z^l$ of layer $l$ and the encoder $V$ and in like manner between the terms $\left(1-Z^l(\mathbf{x})\right)$ and $U$. The final output layer provides the function of acoustic pressure $\netp(t,\rr)$ as given by the neural network at any point in the continuous domain.

The mMLP is trained to fit the measured pressure $\tilde{p}(t_n, \rr_{m})$ as well as fulfil the wave equation so that its objective is delineated by
\begin{alignat}{2}\label{eq:PINN_train}
&\underset{w, \netp}{\arg \min }  \sum_{m \in M} \sum_{n \in N}\left| \left(\netp(t_n, \rr_{m})-\tilde{p}(t_n, \rr_{m})\right)\right| \nonumber \\   
 &\text { s.t. } \nonumber \\
&f\left(t, \rr \right) = \left( \nabla^{2} - \frac{1}{c^{2}} \frac{\partial^2}{\partial t^2} \right) \netp(t, \mathbf{r}) = 0, \,\, \forall \; \mathbf{r} \in \Omega,\; t \in \mathbb{R}^{+}, \nonumber \\
\end{alignat}
%
where $\rr_{m}$ and $t_n$ refer to the positions and time instance of the measured pressure in a room, and $w$ refers to the neural network parameters ($w \subset \mat{W}_l$). The residual function $f(\rr,t)$ applies a constraint on the network to fulfil the wave equation in two dimensions. It is worth noting that the Laplacian $\nabla^2$ is obtained via auto-differentiation, a method used to efficiently calculate derivatives of functions, and is often associated with the underlying mechanism for backpropagation, the fundamental algorithm used in deep learning to train neural networks.

Furthermore, the network uses an adaptive loss function in order to ``learn'' how to balance the data and PDE terms in Eq.\,\eqref{eq:PINN_train}. \cite{self-adaptive-PINN} This loss function can be written as
\begin{equation}\label{eq:adaptive_weights}
\begin{gathered}
L(\varepsilon , w, N_{d}, N_{f})= \frac{1}{2 \varepsilon_d^2}  L_{data}\left(w , N_{d}\right) + \frac{1}{2 \varepsilon_f^2} L_{PDE}\left(w , N_f\right)\\
+\log \varepsilon_d \varepsilon_f,
\end{gathered}
\end{equation}
with 
\begin{equation} \label{eq:loss_terms}
\begin{gathered}
L_{PDE}\left(w , N_f\right)=\frac{1}{N_f} \sum_{n_f=1}^{N_f}\left| f\left(t_{n_f}, \rr_{n_f}; w\right)\right|, \\
L_{\text {data }}\left(w, N_{\text{data}}\right)= \frac{1}{N_d} \sum_{n_d=1}^{N_d}\left| \netp(t_{n_d}, \rr_{n_d}; w)-\tilde{p}(t_{n_d},\rr_{n_d})\right|,
\end{gathered}
\end{equation}
where $N_f, N_d$ are the number of collocation points where the PDE and data are evaluated respectively. The parameters $\varepsilon_f$ and $\varepsilon_d$ are adaptive weights which allow for the network to automatically assign the weights of individual loss terms. This is done by updating these parameters in each iteration based on maximum likelihood estimation.

Once the network is trained, one can obtain the pressure field $\netp(t. \rr) \; \forall \mathbf{r} \in \Omega,\; t \in \mathbb{R}^{+} $ , as well as the particle velocity via Euler's equation of motion (a result of conservation of momentum)
\begin{equation}\label{eq:p_velocity}
    \netu(t, \rr) = -\frac{1}{\rho}\int_{t_0}^{t} \nabla \netp(t, \rr) \diff t,
\end{equation}
where the pressure gradient $ \nabla \netp(t, \rr) $ is obtained with first order auto-differentation. 

Importantly, one can obtain the instantaneous intensity of the sound field in the domain, characterizing how acoustic energy is flowing in the domain of observation, as
\begin{equation}\label{eq:insta_intensity}
    \netI(t, \rr) = \netu(t, \rr) \netp(t, \rr),
\end{equation}
allowing for the full characterisation of the measured sound field.
\section{Methodology}\label{sec:method}
\subsection{Experimental dataset}\label{subsec:experimental_data}
\noindent
The experimental dataset used to train and evaluate the PINN consists of RIRs measured in a dampened room at the Technical University of Denmark (DTU) in Lyngby, Denmark and has been made publicly available. \cite{Karakonstantis2023_Data} The room is used for sound field control experiments and is equipped with panel absorbers on the walls, ceiling, and rear wall. The reverberation time and dimensions of the room can be seen in Tab. \ref{tab:room_char}. The RIRs were measured using a Universal Robots UR5 robotic arm with a 1/2" omnidirectional Br{\"u}el \& Kjær pressure microphone at a grid of 900 positions in the middle of the room. From these 900 positions, only 100 positions were used as training data for the networks, as seen in the left of Fig. \ref{fig:training_data_layout}. The RIRs were obtained using the exponential sine sweep and inverse filter method, which allows for a high signal-to-noise ratio, given an adequate sweep length, while using a Dynaudio BM6 2-way passive loudspeaker as a source. The layout of the complete dataset, as well as a panoramic view of the robot arm in the room can be seen in Fig. \ref{fig:complete_dataset_experimental}.
\newcommand{\captionone}{(Color online) Measurement configuration in the room at DTU showing the respective loudspeaker positions, as well as the measurement aperture (top) and a panoramic view of the robot arm in the measurement room at DTU (bottom).}
\ifdefined\showfigures
\begin{figure}[!t]
\centering
\includegraphics[width=\reprintcolumnwidth]{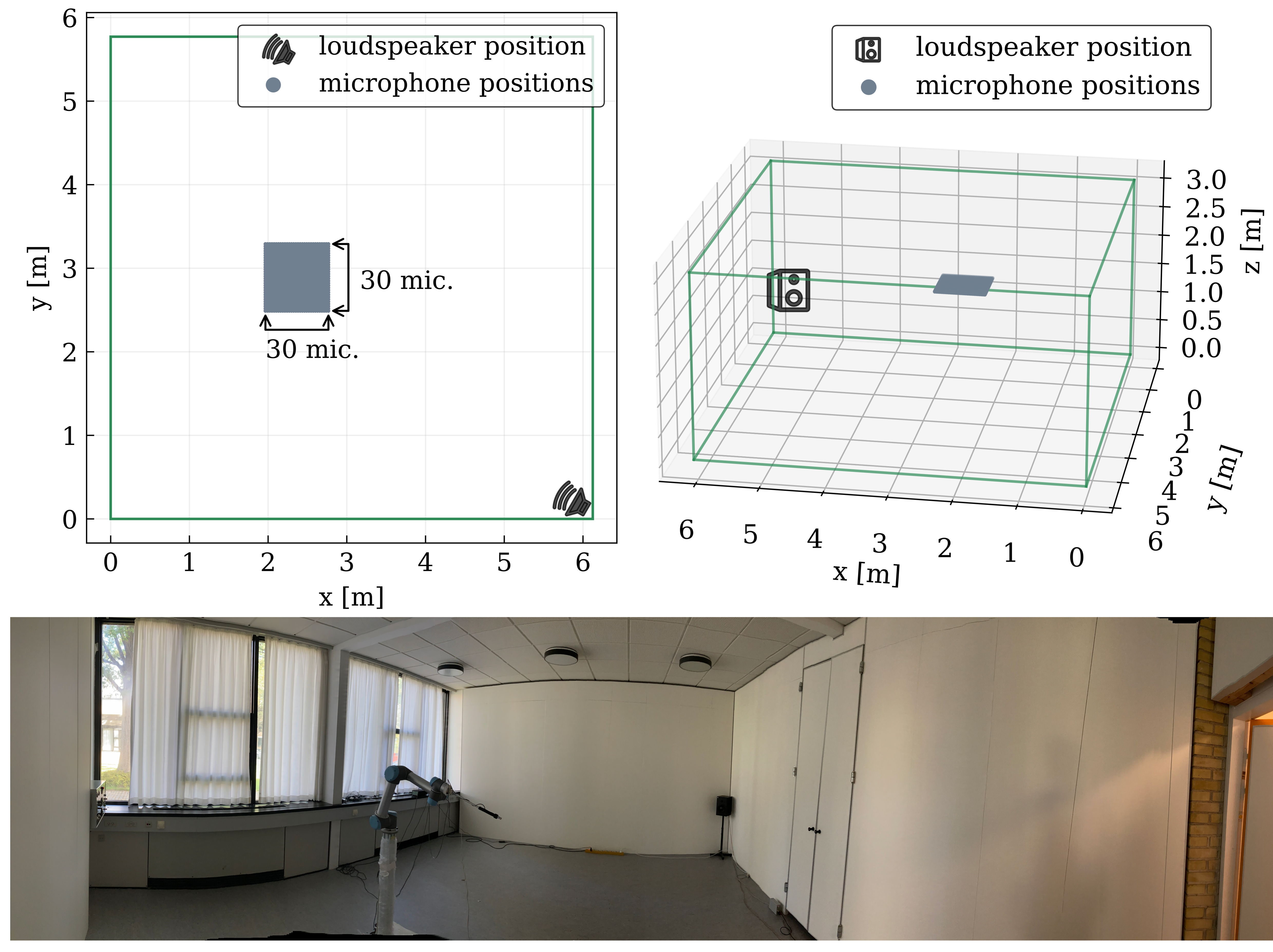}
\caption{\captionone}
\label{fig:complete_dataset_experimental}
\end{figure}
%
\begin{table}[!t]
\renewcommand{\arraystretch}{1.3}
\caption{Size and reverberation times in octave bands from the studied room (DTU)}
\label{tab:room_char}
\centering
\resizebox{\reprintcolumnwidth}{!}{%
\begin{tabular}{|c|cccccc|}
\hline
Dimensions (m$^3$)                                     & \multicolumn{6}{c|}{Reverberation Time - $T_{60}$ (s)}                                                                                                                  \\ \hline
\multirow{2}{*}{$6.12\times 5.77 \times 3.07$} & \multicolumn{1}{c|}{125 Hz} & \multicolumn{1}{c|}{250 Hz} & \multicolumn{1}{c|}{500 Hz} & \multicolumn{1}{c|}{1 kHz} & \multicolumn{1}{c|}{2 kHz} & 4 kHz \\ \cline{2-7} 
                                               & \multicolumn{1}{c|}{0.5}    & \multicolumn{1}{c|}{0.43}   & \multicolumn{1}{c|}{0.37}   & \multicolumn{1}{c|}{0.52}  & \multicolumn{1}{c|}{0.52}  & 0.44  \\ \hline
\end{tabular}
}
\end{table}
%
\newcommand{\captiontwo}{(Color online) Layout of PINN training data for experimental dataset (left - 100 receiver points) and simulated dataset (right - 961 receiver points).}
\ifdefined\showfigures
\begin{figure}[!t]
\centering
\includegraphics[width=\reprintcolumnwidth]{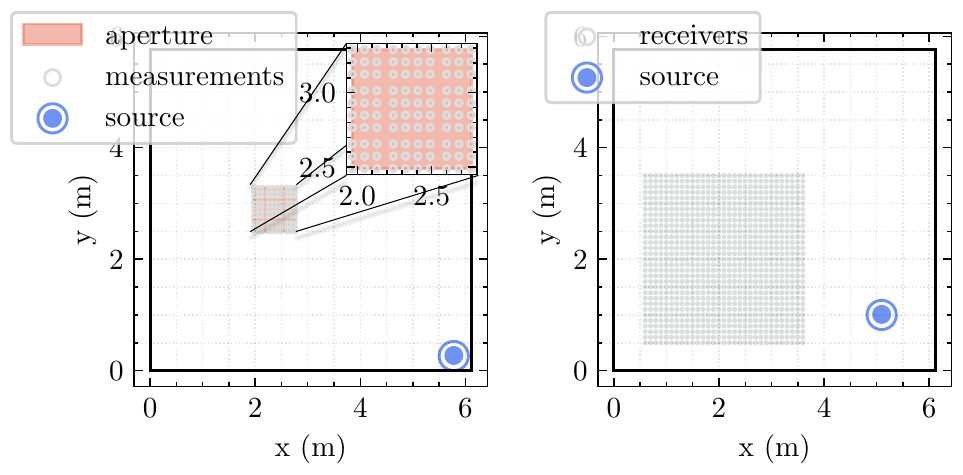}
\caption{\captiontwo}
\label{fig:training_data_layout}
\end{figure}
%
\subsection{Simulated dataset}
\noindent
A model of the same dampened room was created in the room acoustic simulation program Odeon. \cite{naylor1993odeon} The material properties were adjusted so that similar room acoustic parameters could be derived from the numerical model and the measured impulse responses ($T_{20}$, $C_{50}$, $D_{50}$). From this model, we extracted simulated impulse responses using a horizontal grid of $61 \times 61$ receivers (a total of 3721 receivers), spaced 5 cm apart, covering a $3 \times 3$ m$^2$ aperture. Within this dataset, we selected responses that were simulated at intervals of 10 cm from each other, resulting in a subset of $31 \times 31$ positions (a total of 961 receivers) reserved specifically for training, as depicted on the right side of Fig. \ref{fig:training_data_layout}. The remaining data is exclusively used for comparison with the PINN reconstructed responses. Regarding the simulation, the sound impulse originates from an omnidirectional source situated at about 1.5\,m from the grid of receivers. The simulated dataset covers a larger space than the measurements, as the grid of receivers is a square of side dimension 3\,m.

Odeon is a hybrid geometrical acoustics program, in which the early part of the impulse response is obtained with an image source method and the late part is calculated with a ray-tracing algorithm. In this study, the transition order is set to 3, meaning that for each reflected ray, the three first reflections are treated with the image source method. Due to the high wall absorption in the room, these early reflections are expected to dominate in the impulse responses.

The Odeon algorithm is energy-based, so the actual output is the squared impulse response at a given receiver position. For auralisation purposes, Odeon adds a random sign to each reflection in order to generate a pressure impulse response. Therefore, the pressure impulse responses simulated by Odeon are closely related, but not identical, to the physical room measurements. Nevertheless, the random signs are kept consistent from receiver to receiver, for each reflection. The relative levels and times of arrival of each reflection should be well reproduced in the simulation.

\subsection{Architecture and training of the Physics-Informed Neural Network}
\noindent
For the non-linear activation functions, we use sinusoidal activation functions because they have been shown to be effective in neural networks as universal approximators when initialised correctly. \cite{sitzmann2020implicit} Sinusoidal activations are particularly useful for modelling high frequency or periodic data, or data structures that require higher order spatial derivatives, which other activation functions may not be able to handle. Benbarka et al. \cite{SIREN_Fourier_Series} provide further information on how MLPs with sinusoidal activation functions can be equivalent to d-dimensional Fourier mappings, with the weights of the perceptron corresponding to a Fourier series. The sinusoidal activation functions receive a hyperparameter $\omega_0$ such that $\sigma\left( \vec{x} \right) = \sin\left( \omega_0 \vec{x} \right)$, which controls the frequency mappings of the Fourier series. In our experiments, we initially set $\omega_0 = 15$ for the sinusoidal activation functions, which is within a common range present in the literature. To investigate its impact, we conducted a limited parameter study ranging from $\omega_0 = 5$ to $\omega_0 = 30$. The hyperparameter $\omega_0$ controls the frequency of features captured by the network. Smaller values prioritise low-frequency patterns, while larger values emphasise finer details. Despite these differences, all values eventually converged to similar results.

Moreover, we use the Adam optimiser for both the adaptive weights and the neural network parameters, with learning rates of $\eta_\varepsilon = 2\cdot10^{-4} $ and $\eta_w = 2\cdot10^{-5} $ respectively. The parameters regarding the number of collocation points for each loss term are set to  $N_f = N_d =25000$, with points selected using Latin hypercube sampling in both space and time for the PDE term, and the training data (i.e. experimental or simulated) was sampled at uniformly random spatio-temporal time instances for the data term, also corresponding to a batch-size of $N_{\text{data}} = 25000$. Additionally, the parameters controlling the weights of the individual loss terms in Eq.\,\eqref{eq:adaptive_weights} were initialised to $\varepsilon_d = 1$ and $\varepsilon_f = 10$ respectively, to prioritise the fidelity on the actual data.

In the case of the simulated data, both the MLP and modified MLP network architectures employ 800 neurons in each layer, except for the input layer with 3 neurons and the output layer with 1 neuron. On the other hand, the experimental data demonstrates satisfactory performance with a reduced number of neurons, utilising only 512 neurons with the modified MLP network architecture. The training process comprises $150000$ iterations, conducted on a single 32 GB NVIDIA V100 GPU, which takes approximately 9 hours for the simulated data and 6 hours for the experimental data, with memory utilisation restricted to 25\% of the total capacity.

Figure \ref{fig:loss_terms} provides a visual representation of the network's training progress over $150000$ iterations, including validation. The bottom row of the figure illustrates the convergence behavior of the Mean Absolute Error (MAE) for two loss terms (Eq.\,\eqref{eq:loss_terms}): the PDE term, characterising the network's fidelity to the wave equation, and the data term computed over the 100 fitted positions, representing the agreement with experimental data. Both loss terms exhibit a substantial initial decrease, suggesting rapid model adaptation. Notably, after approximately $20000$ iterations, the MAE of the PDE term stabilises, whereas the data term continues to exhibit a decline. The validation includes the MAE over $5$ additional positions (not included in the training set) the pressure values of which are sampled at random time instances during training. The validation loss also reaches a stable state around $20000$ iterations, implying that the reconstruction quality is poised to benefit from additional data. However, it should be noted that the although the data loss and validation loss curves start to diverge after $20000$ iterations, the interpolation capabilities of the network are not compromised as is common with other types of neural networks (lack of generalisability). The top row of the figure presents the adaptive weights assigned to these loss terms during training (Eq.\,\eqref{eq:adaptive_weights}). The weight assigned to the PDE term exhibits a consistent, monotonic increase. In contrast, the data term weight demonstrates a sharp ascent up to $20000$ iterations, followed by a more gradual increase, highlighting the network's emphasis on fitting the experimental data.

\newcommand{\captionthree}{(Color online) Convergence and adaptive weighting analysis during 150000 iterations of the PINN trained on experimental data.}
\ifdefined\showfigures
\begin{figure}[!t]
\centering
\includegraphics[width=\reprintcolumnwidth]{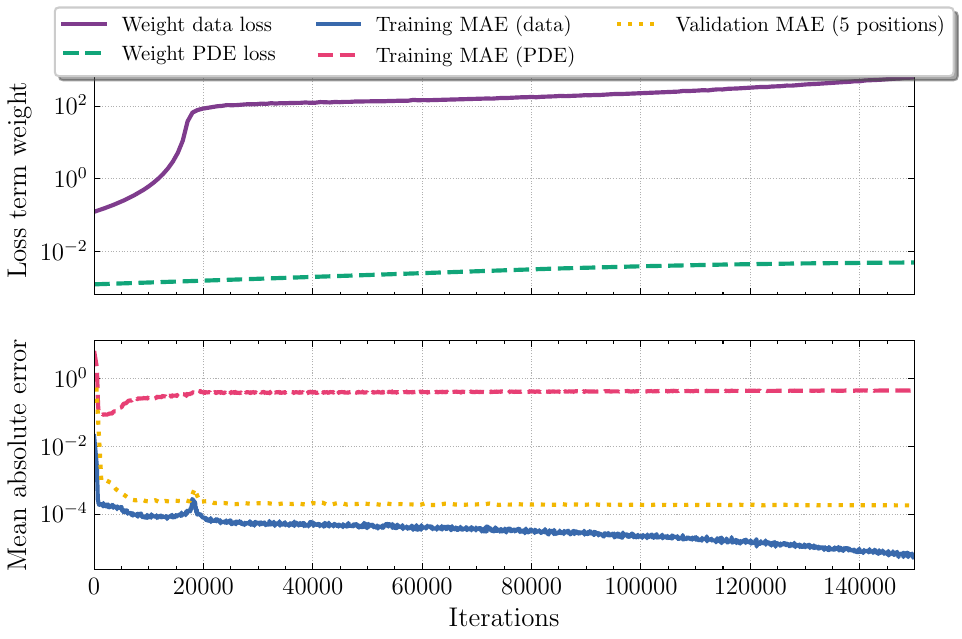}
\caption{\captionthree}
\label{fig:loss_terms}
\end{figure}

\subsection{Alternative approaches for sound field reconstruction: Data-driven neural network and wave-based regression}\label{subsec:wave_regression}
\noindent
In this study, we consider three relevant baseline methods for comparison: A Deep Image Prior (DIP) or Deep Prior approach\cite{pezzoli2022deep} and two wave-based regression methods using sparsity promoting spherical waves\cite{caviedes2023spatio,antonello2017room}and plane waves \cite{verburg2018reconstruction}. 

Deep Prior is based on image inpainting where the RIRs measured $\tilde{\P} \in\R^{N\times M}$ are considered an incomplete image
\begin{equation}\label{eq:deep_prior}
\tilde{\P} = \P\mathbf{S},  
\end{equation}
where $\P \in\R^{N\times M}$ is the sound field to recover and $\mathbf{S}\in\R^{N\times M}$ is a binary matrix that activates or deactivates pixels depending on the availability in the measured data. The sound field $\P \in\R^{N\times M}$ is approximated by training a convolutional neural network $\mathcal{NN}(\pmb{\theta}, \mathbf{Z})$ under the following optimization problem
\begin{equation}
    \pmb{\hat{\theta}} = \argmin_{\pmb{\theta}}\E[\mathcal{NN}(\pmb{\theta}, \mathbf{Z})\mathbf{S}-\tilde{\P}].
\end{equation}
where $\pmb{\theta}$ are the neural network parameters, and $\mathbf{Z}$ is a fixed, latent-space vector initialised to be distributed as a standard normal variable.

As the experimental and simulated datasets consist of 2D pressure apertures, instead of employing a linear array of measured RIRs for the Deep Prior fitting, as outlined in \cite{pezzoli2022deep}, we adopt multiple 2D pressure snapshots. The binary matrix defined by Eq.\,\eqref{eq:deep_prior} masks the complete set, ensuring equivalence of datasets for both PINN and DIP.

The wave-based regression methods involve a linear combination of basis functions, described by Eqs.\,\eqref{eq:convolution} and \eqref{eq:freq_domain_basis}. The first method employs spherical waves $\varphi_l(t, \mathbf{r})$ as basis functions, given by
\begin{equation}
    \varphi_l(t, \rr) = \frac{1}{4\pi d_l}\delta\left(t-\frac{d_l}{c}\right),
\end{equation}
where  $d_l=\|\rr-\rr_l\|_2$ and the Dirac function $\delta\left(\cdot\right)$ is modelled as a fractional delay filter. If discretized, the convolution in Eq.\,\eqref{eq:convolution} can be made explicit as a double summation
\begin{equation}\label{eq:explicit_convolution}
    p[n, m] \simeq \sum_{l=1}^{L}\sum_{i=1}^{N} \varphi_{l}[n-i, m]\alpha_l[i].
\end{equation}
Reference \cite{antonello2017room} presents this method in detail.

The second method, outlined by Eq.\,\eqref{eq:freq_domain_basis}, assumes plane wave propagation so that the basis functions take the form
\begin{equation}\label{eq:planewaves}
\psi_{l}(\omega, \rr) = \exp\left(-j \frac{\omega}{c} \rr\right),
\end{equation}
leading to the task of recovering the coefficients $\beta_{l}(\omega)$. 
The unknown time domain coefficients $\alpha_l[i]$ and are calculated via Bayesian optimization and variational inference, with a Laplace prior to promote sparse solutions, following the methods in, \cite{caviedes2021gaussian} whereas the frequency domain coefficients were estimated using a regularised least squares framework, also using a sparse prior. \cite{verburg2018reconstruction}

\section{Results}\label{sec:Results}

\subsection{Sound field characterisation of experimental data}
\noindent
Similar to holography methods, the PINN is able to reconstruct the complete field, namely the pressure field, as well as the particle velocity and instantaneous intensity using the analytical expressions of Eqs. \eqref{eq:p_velocity} and \eqref{eq:insta_intensity} respectively.

Figure \ref{fig:sf_characterisation} shows the reconstructed pressure (top row), particle velocity (middle row) and instantaneous intensity (bottom row), using only the 100 measurements of the complete experimental dataset. This encompasses various time snapshots, ranging from the direct and early reflections (e.g., 11 ms, 20 ms, 51 ms) to the later, more diffuse part of the sound field (e.g., 82 ms). At 11 ms, clear identification of the direct wavefront followed by the ground reflection is evident. Remarkably, similar to the outcomes observed in the simulated dataset, the PINN demonstrates a faithful and comprehensive reconstruction of the wavefronts, even with the measurements being approximately 9 cm apart.
\newcommand{\captionfour}{(Color online) Pressure (top) particle velocity (middle) and instantaneous intensity (bottom) of the reconstructed sound field using 100 measurements.}
\ifdefined\showfigures
\begin{figure*}[ht!]
\centering
\includegraphics[trim=0.18cm 0.17cm 0.16cm 0.11cm, clip,width=\linewidth]{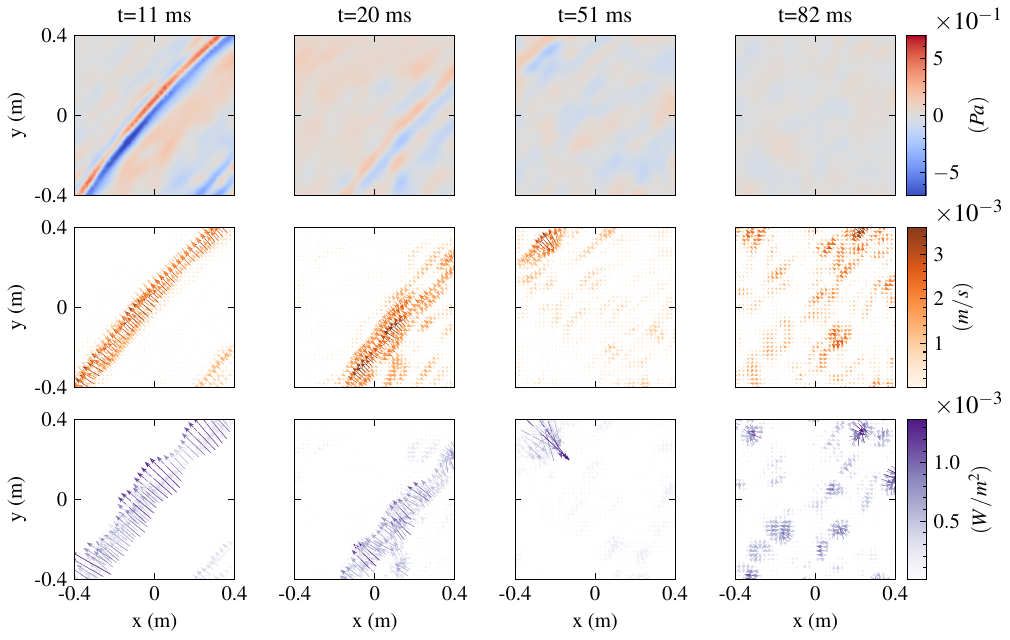}
\caption{\captionfour}
\label{fig:sf_characterisation}
\end{figure*}
Due to the proportional relationship between velocity and the gradient of sound pressure, regions with high values (characterized by longer arrowheads or darker colour) signify rapid spatial changes in sound pressure. As anticipated, such characteristics are evident in the early reflections of the snapshots at time t $= 11, 20,$ and $51$ ms. Conversely, at t $= 82$ ms, the sound energy initiates a more isotropic spatial distribution, indicating the emergence of a diffuse sound field. The instantaneous intensity is the primary indicator of distinct wavefronts, clearly demonstrating distinct waves at times t $= 11, 20,$ and $51$ ms. This does not become immediately evident from the pressure alone. For example, the intensity at t$= 51$ ms indicates a lateral reflection from the back of the room, whereas the wave direction is ambiguous when examining the top row consisted of scalar pressure values. Finally, at 82 ms the instantaneous intensity field exhibits distinct regions characterised by converging intensity flows. These patterns are to be expected due to the presence of reactive components in the intensity, resulting from the interactions of waves. The variability in phase lag between pressure and velocity contributes to this phenomenon, leading to a departure from the earlier reflections where wavefronts remained discernible, and waves were locally planar with pressure and velocity being approximately in phase.
 \subsection{Benchmarking training strategies for sound field reconstruction}
\noindent
For evaluation of the reconstructed sound fields we use Pearson's correlation coefficient defined as
\begin{equation}
    \rho(t, \rr) = \frac{\E[p(t, \rr)\hat{p}(t, \rr)]-\E[p(t, \rr)]\E[\hat{p}(t, \rr)]}{\sqrt{\E[p^2(t, \rr)]\E[p(t, \rr)]^2}\sqrt{\E[\hat{p}^2(t, \rr)] - \E[\hat{p}(t, \rr)]^2}},
\end{equation}
between any true (reference) $p(t, \rr)$ and reconstructed $\hat{p}(t, \rr)$ pressure along any axis of the pressure (time or space). This is a simple measure, showing how the reconstructed room reflections might covary with the experimental truth. It ranges between -1 (perfectly negatively correlated), through 0 (no linear correlation), to 1 (perfectly positively correlated).

Furthermore, we find the root-mean-square error (RMSE) a good measure of discrepancy between the experimental truth and the reconstructed pressure. This can be defined by
\begin{equation}\label{eq:RMSE}
    \text{RMSE} = 10\log_{10}\left(\sqrt{\E\left[ \left(p(t, \rr) - \hat{p}(t, \rr) \right)^2\right]}\right).
\end{equation}
In many cases, it's more preferable to employ the normalised RMSE. This involves normalising the quantity within the logarithm in Eq.\,\eqref{eq:RMSE} by the mean of the square root of the energy in the signal (i.e., $\sqrt{\E\left[p(t, \rr)^2\right]}$). This normalised metric represents the error as a ratio relative to the magnitude of the ground truth, with lower values indicating better agreement between the estimated and ground truth values.

To comprehensively assess the capabilities of the choice of neural network in reconstructing the various acoustic field quantities, we demonstrate the performance of three distinct networks. These networks comprise a modified multi-layer perceptron (mMLP), an mMLP incorporating a physics loss based on the wave equation (mMLP-PINN), and a simple MLP with a physics loss (MLP-PINN). Figure \ref{fig:Quantitative_model_selection} shows the RMSE and correlation between the ground truth pressure fields and the reconstructed fields obtained from each network in steps of 10 ms. Additionally, it's important to note that these results were obtained by fitting the whole dataset of 900 experimental RIRs purely for the purpose of model selection. The results revealed that the mMLP-PINN network displayed the most notable performance, achieving significantly higher correlation compared to the other two trained networks. Subsequently, the mMLP-PINN demonstrated an approximately 10 dB reduction in RMSE when compared to the other models. 

The inclusion of the physics loss in the MLP-PINN seemingly contributed to better regularisation of the solution, leading to slightly improved performance compared to the mMLP without physics loss. However, an important observation emerged when examining the effects of the physics loss on the pressure field. At time t$=0$ s, where the pressure is theoretically zero, the MLP-PINN showed the lowest correlation coefficient, indicating the introduction of negligible noise to the pressure field. Similarly, the mMLP-PINN exhibited a slightly worse correlation at time 0 than the mMLP, further suggesting potential effects of the physics loss on the pressure field. These findings provide valuable insights into the advantages and limitations of each architecture and the impact of physics-based losses in acoustic field reconstruction tasks.
\newcommand{\captionfive}{(Color online) Quantitative comparison of reconstruction performance between a modified MLP without physics loss (mMLP), a mMLP with physics loss (mMLP-PINN) and a MLP with physics loss (MLP-PINN).}
\ifdefined\showfigures
\begin{figure}[!t]
\centering
\includegraphics[width=\reprintcolumnwidth]{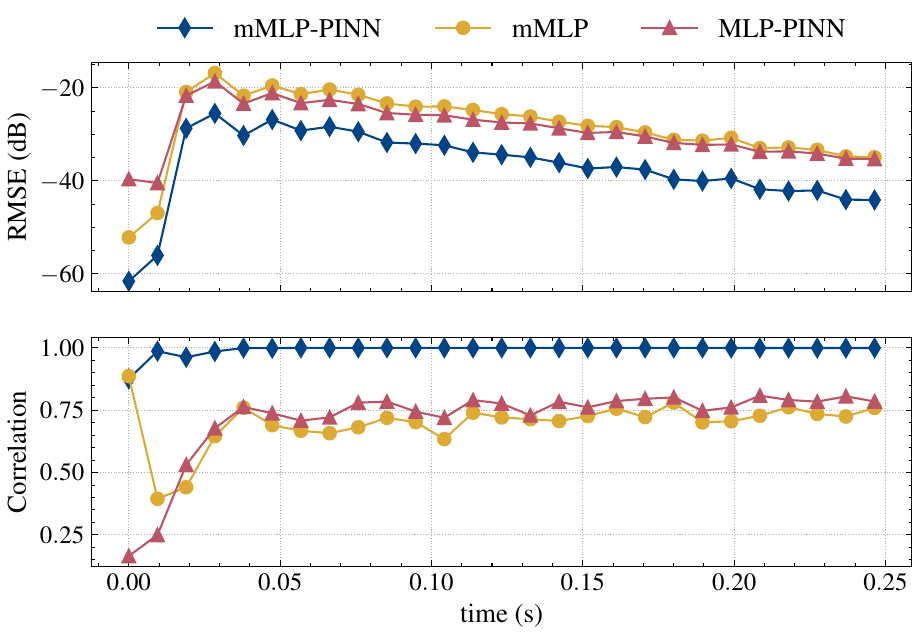}
\caption{\captionfive}
\label{fig:Quantitative_model_selection}
\end{figure}

\subsection{Gridless sound field mapping in simulated environments}
\noindent
We examine the use of PINNs as a tool to enhance the output of acoustic simulation methods, as they make it possible to provide a fast (almost instantaneous) grid-less map of the sound field in a domain. Put differently, one can numerically obtain a sound field across a dense spatial grid and subsequently train a PINN on this data. This enables the network to provide an efficient and continuous spatial mapping.

Leveraging this capability of PINNs to operate without predefined grids, we are able to visualise a refined version of the simulated pressure field to assess performance of the method. The top row of Fig. \ref{fig:Odeon_sf_snapshots} illustrates the simulated pressure fields at different time snapshots (ranging from 13ms to 41ms), offering insights into the temporal evolution of the sound field. Below, we present the corresponding super-resolved PINN reconstructions, highlighting the model's ability to faithfully reproduce the pressure distribution across the spatial domain in a higher resolution. The PINN provides refined wavefronts, without discontinuities, which becomes especially apparent when comparing the data at $t = 13$ ms to the PINN reconstruction at the same time instance.  

Figure \ref{fig:Interp_RIRs_odeon} displays a magnified view of the $3\times3$ grid of positions employed for training the PINN (circle-shaped markers) and examples of reconstruction positions (interpolated - indicated by the `x'-shaped markers). The figure further illustrates the reconstructed RIRs superimposed on the true RIRs, accompanied by a quantitative assessment of reconstruction quality, measured in terms of correlation and linear scale RMSE. The noteworthy performance of the PINN in reconstructing pressure fields demonstrates its ability to accurately capture sound field attributes at positions not previously sampled or simulated. These findings underscore the effectiveness of PINNs for gridless sound field mapping in simulated environments, where inference time remains noticeably efficient, thus opening avenues for streamlined auditory navigation in 6DOF virtual environments, particularly beneficial in scenarios where constraints in free movement can hinder localisation.\cite{lokki2000case}

%
%
\newcommand{\captionsix}{(Color online) Snapshots of the simulated sound field (left) and the PINN super-resolved sound field (right) with the simulated data.}
\ifdefined\showfigures
\begin{figure}[!t]
\centering
\includegraphics[width=\reprintcolumnwidth]{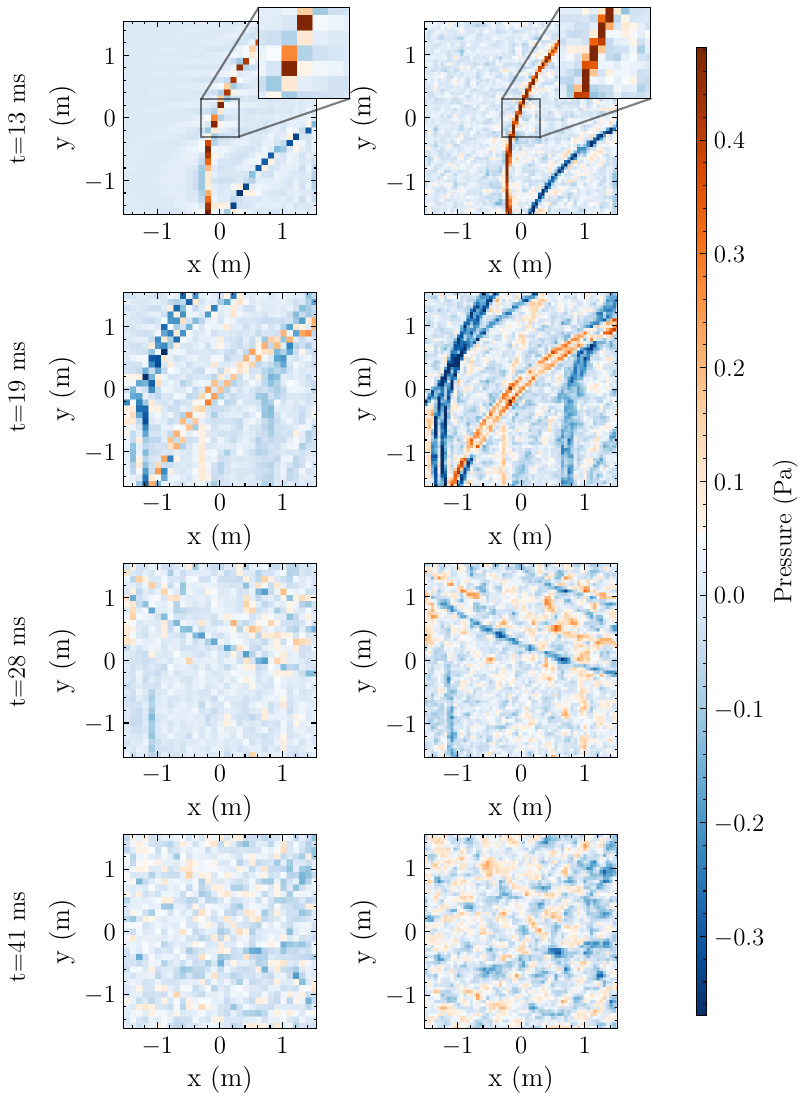}
\caption{\captionsix}
\label{fig:Odeon_sf_snapshots}
\end{figure}
%
\newcommand{\captionseven}{(Color online) Simulated data positions (indicated by blue circle markers), alongside positions of the interpolated RIR positions (marked with pink 'x' symbols). The reconstructed RIRs are superimposed upon the true RIRs for comparison.}
\ifdefined\showfigures
\begin{figure}[!t]
\centering
\includegraphics[trim=0.02cm 0.16cm 0.14cm 0.13cm, clip,width=\reprintcolumnwidth]{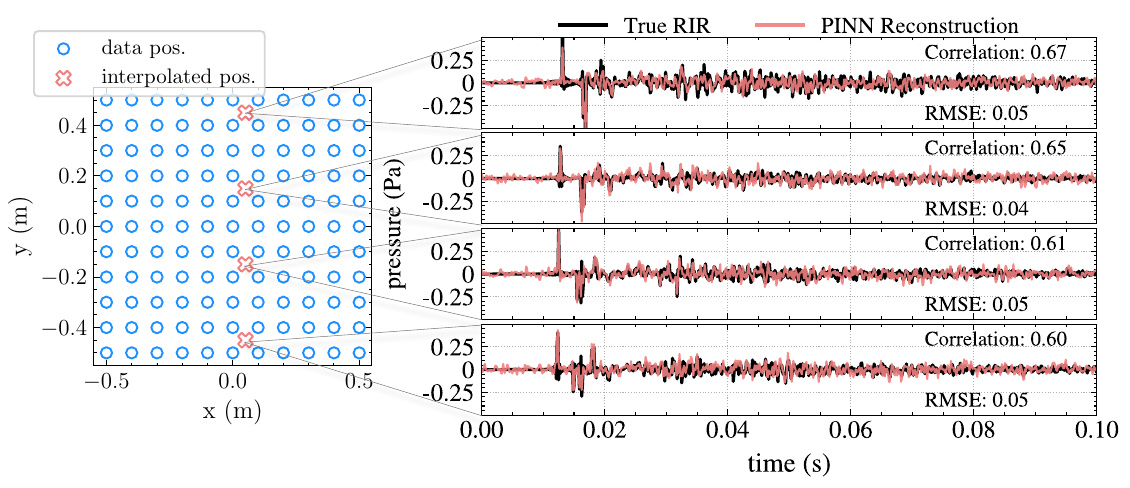}
\caption{\captionseven}
\label{fig:Interp_RIRs_odeon}
\end{figure}
%
%
\section{Comparison with state-of-the-art methods for sound field reconstruction using experimental data}
\noindent
Figure \ref{fig:method_comparison} shows a comparison of the modified Deep Prior network (DIP) estimated pressure, and the PINN estimated pressure. For the DIP the data is a discretised pressure ``image'' with the same size as the aperture, but masked with zeros in positions without available measurements (see Fig. \ref{fig:training_data_layout} - left). For purposes of brevity, we examine three snapshots (t $= 13, 21,$ and $39$ ms). These three snapshots were selected on the basis that they contain the highest changes in energy density over the medium, and the reconstructed pressure should conform to homogeneous wavefronts. One can see that the DIP often fails to properly capture the continuous variation in pressure, as opposed to the PINN which, although not perfect, still manages to resemble the dominating waves of the true pressure field. This becomes evident in the absolute pressure-wise difference in the bottom row of Fig. \ref{fig:method_comparison}, where the PINN displays a higher accuracy. It should be noted that, albeit the rapid training and inference of the DIP method, in order to reconstruct RIRs with this spatial configuration, one would require to train the network as many times as there are time samples in the RIRs, which is not optimal.
\newcommand{\captioneight}{(Color online) Comparison of data-driven reconstructed pressure snapshots (DIP) with the proposed (PINN) reconstructed pressure snapshots.}
\ifdefined\showfigures
\begin{figure*}[ht!]
    \centering
    \includegraphics[trim=0.02in 0.02in 0.03in 0in, clip,width=\linewidth]{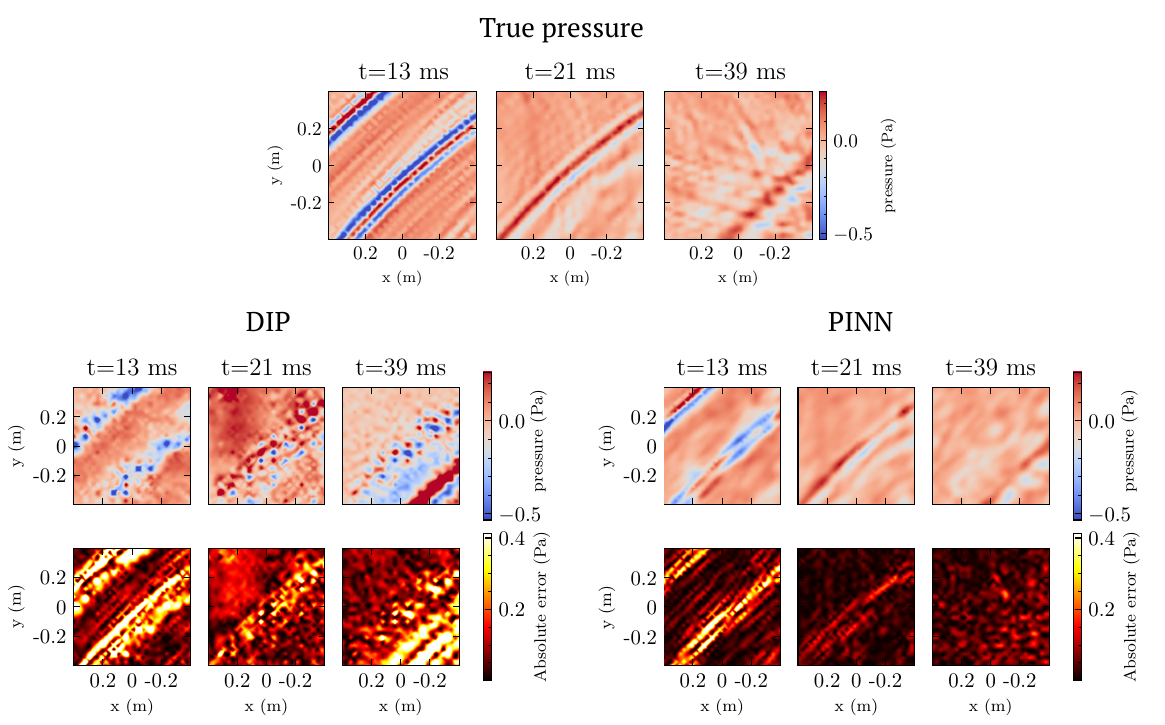}
    \caption{\captioneight}
    \label{fig:method_comparison}
\end{figure*}

We investigated how the PINN fares in comparison to elementary wave basis reconstruction, as outlined in \ref{subsec:wave_regression}. For clarity, we define the abbreviations and names given to these baseline models, where the time domain (TD) spherical wave using a Laplace (sparse) prior is denoted as `TD Laplace Prior', whereas the frequency domain plane wave (PW) model, using compressive sensing and regularised least squares (RLS), is termed `PW expansion RLS'. 

Figure \ref{fig:sf_reconstruction_comparison} provides an assessment of the reconstructed pressure for both wave-based regression models and the PINN. Both spatial normalised RMSE and correlation are evaluated in snapshot intervals of 1.8 ms. This allows us to evaluate the pressure variations in spatial windows with slight overlap. In case of the PINN, the normalised RMSE is approximately $2$ dB less than the other two models for most of the early part of the spatio-temporal response (between $0.01$ and $0.03$ seconds). This suggests that the PINN responses are less prone to (Gibbs phenomena) pre-ringing artefacts,\cite{gibbs1898fourier} which are common in wave-based methods when an insufficient number of basis functions (i.e., truncated series expansion) are available to represent the sound field. The same can be deduced when examining the correlation between predictions and truth of each of the models, also during the same time interval. The PINN is much more correlated to the true sound field, being able to properly model the amplitudes and phases of the direct response and early reflections. The stronger performance of wave-based methods in the later sound field ($0.06-0.09$ ms) can be attributed to the more isotropic and random behavior of reverberant sound fields, which aligns with random wave field theory. \cite{jacobsen2013fundamentals} Thus, PINNs may face challenges in capturing the complex interactions within this region, emphasising the need to match the modeling approach with the acoustic characteristics of the environment.
\newcommand{\captionnine}{(Color online) RMSE (top) and Correlation (bottom) of reconstructed pressure (snapshots) over space for the TD Laplace prior, PW expansion RLS and PINN models.}
\ifdefined\showfigures
\begin{figure}[!ht]
\centering
\includegraphics[trim=0.02in 0.02in 0.03in 0in, clip,width=\reprintcolumnwidth]{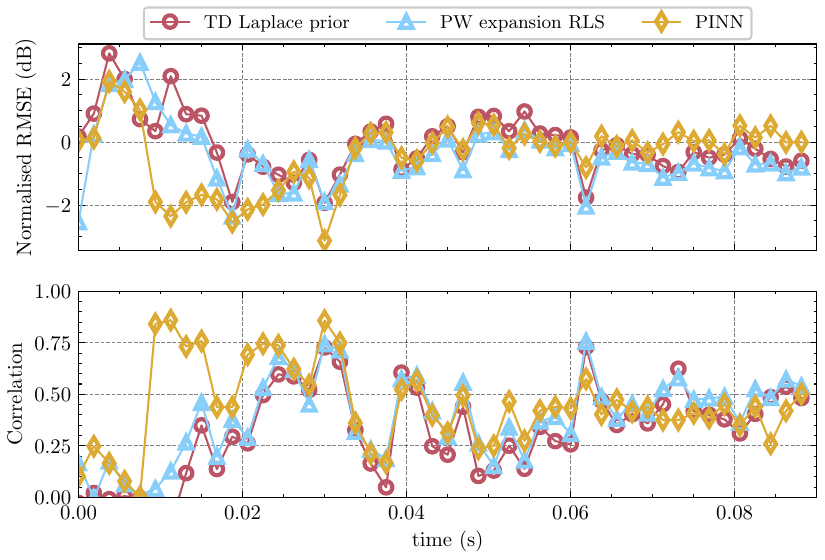}
\caption{\captionnine}
\label{fig:sf_reconstruction_comparison}
\end{figure}

Analysing the reconstructions of single RIRs with respect to experimental truth at varying distances from the center of the planar aperture sheds light on the PINNs ability to extrapolate. These insights are visually represented in Fig. \ref{fig:extrapolation_results}. In this analysis, the PINN was trained with a reduced dataset, featuring $8\times8$ microphones instead of the original $10\times10$, and a smaller aperture size ($60\times60$ cm$^2$ compared to the initial $80\times80$ cm$^2$), deliberately designed to explore the network's ability to extrapolate beyond training data boundaries. The resulting figure shows that as we move further from a data point (measured RIR), the PINN's performance, both in terms of correlation and RMSE, gradually diminishes. This decline in performance as we extrapolate from the training data is a natural consequence, due to the network's dependence on available data points for accurate predictions. Notably, the PINN performs well when closely aligned with a data point. This highlights the significance of having a substantial number of measurenents to ensure dependable reconstruction.

\newcommand{\captionten}{(Color online) Corellation and RMSE of single RIRs along the $y = 10$ cm axis, for a model trained with a $60\times60$ cm$^2$ aperture corresponding to $8\times8$ measured RIRs.}
\ifdefined\showfigures
\begin{figure}[!t]
\centering
\includegraphics[trim=0.1cm 0.16cm 0.1cm 0.11cm, clip,width=\linewidth]{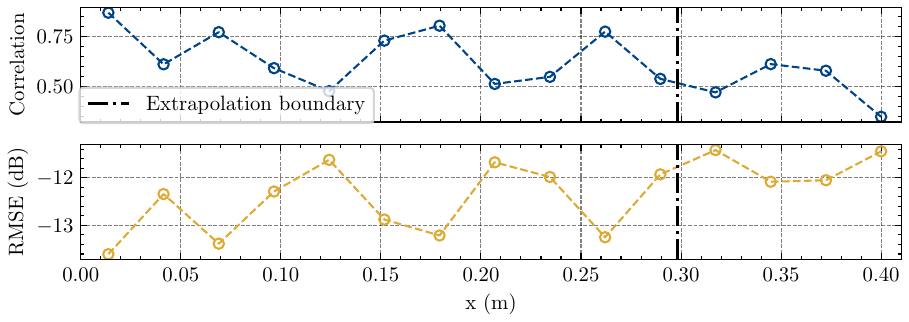}
\caption{\captionten}
\label{fig:extrapolation_results}
\end{figure}

An example of single RIR reconstruction is shown in Fig. \ref{fig:rir_reconstruction_comparison}. As previously observed, the PINN outperforms both methods in terms of error (RMSE) and similarity (correlation) as a function of time, denoted over each of the reconstructed RIRs. This amplifies the notion of PINN suitability for auralising sampled sound fields, since the temporal fine structure (TFS) of the PINN RIR is clearly closer to the true underlying RIR. A poor reconstruction in terms of TFS can severly hinder the inter-aural time differences of an acoustic scene, and consequently, the localisation of a source in a room. \cite{plack2018sense}
\newcommand{\captioneleven}{(Color online) Comparison of RIR reconstruction between the TD Laplace prior (top) and PW expansion RLS (middle) models, with the PINN model.}
\ifdefined\showfigures
\begin{figure*}[!t]
\centering
\includegraphics[trim=0.18cm 0.16cm 0.14cm 0.13cm, clip,width=\linewidth]{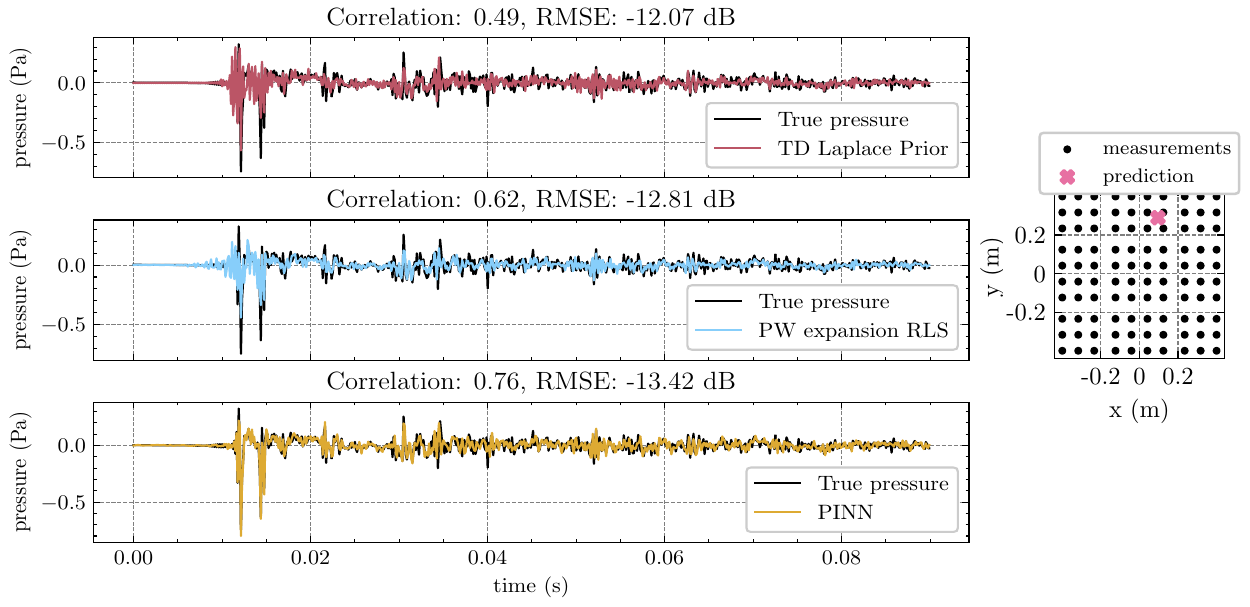}
\caption{\captioneleven}
\label{fig:rir_reconstruction_comparison}
\end{figure*}
%
\section{Conclusions}\label{sec:conclusions}
\noindent
In this paper, the application of PINNs for sound field reconstruction is examined, both with experimental measurements and room acoustic simulations. It is shown that PINNs can be used for complete characterisation of sound fields, reconstructing valuable time-domain field quantities such as particle velocity and instantaneous intensity, despite the challenges posed by experimental conditions. The gridless representation capabilities of PINNs to model sound pressure enables real-time auralisation from any desired position within a predefined aperture, which can lead to improved 6DOF audio applications. The experimental results revealed that PINNs offer advantages over traditional wave-based regression techniques as well as data-driven priors in the form of neural networks. Specifically, PINNs exhibited a high degree of accuracy in reconstructing the early part of room impulse responses, surpassing the performance of the other models in this regard.
In conclusion, the study highlights the applicability of PINNs as a reliable tool for sound field reconstruction tasks, offering versatility in both simulated and real-world acoustic settings. As further research and advancements are made in the field of deep learning, PINNs emerge as a promising avenue for advancing sound field reconstruction, analysis, and reproduction methods. These networks promise to enhance the creation of immersive auditory experiences, aligning with the growing interest in this transformative domain.

\begin{acknowledgments}
This work was supported by the VILLUM Foundation, under grant number 19179, ``Large-scale acoustic holography.'' The authors would like to thank Antonio Andr\'es Figeroa Dur\'an, Franz Heuchel and Manuel Hahmann for the valuable discussions during the conception of this work.
\end{acknowledgments}












\ifdefined\appendcaptions
\section*{Figure captions}
\renewcommand{\labelenumi}{Figure \arabic{enumi}:}
\begin{enumerate}
\setlength{\itemindent}{5ex}
\item{\captionone}
\item{\captiontwo}
\item{\captionthree}
\item{\captionfour}
\item{\captionfive}
\item{\captionsix}
\item{\captionseven}
\item{\captioneight}
\item{\captionnine}
\item{\captionten}
\item{\captioneleven}
\end{enumerate}
\fi

\end{document}